\documentclass[aps,prl,twocolumn,showpacs,groupedaddress]{revtex4}

\usepackage{graphicx}
\usepackage{amsmath}
\usepackage{amssymb}
\usepackage{bm}
\usepackage{color}
\usepackage[colorlinks=true,citecolor=blue,urlcolor=blue]{hyperref}
\usepackage{array}
\usepackage{pdfpages}
\begin{document}
\title{Spin 1 condensates at thermal equilibrium : a $SU(3)$ coherent state approach}
\author{V. Corre, T. Zibold, C. Frapolli, L. Shao, J. Dalibard and F. Gerbier\email{fabrice.gerbier@lkb.ens.fr}} 
\affiliation{Laboratoire Kastler Brossel, Coll\`ege de France, CNRS, ENS-PSL Research
University, UPMC-Sorbonne Universit\'es, 11 place Marcelin Berthelot, 75005 Paris}
\date{\today}
\begin{abstract}
We propose a theoretical framework based on $SU(3)$ coherent states as a convenient tool to describe the collective state of a Bose-Einstein condensate of spin 1 atoms at thermal equilibrium. We work within the single-mode approximation, which assumes that all atoms condense in the same spatial mode. In this system, the magnetization $m_z$ is conserved to a very good approximation. This conservation law is included by introducing a prior distribution for $m_z$ and constructing a generalized statistical ensemble that preserves its first moments.   In the limit of large particle numbers, we construct the partition function at thermal equilibrium and use it to compute various quantities of experimental interest, such as the probability distribution function and moments of the population in each Zeeman state. When $N$ is large but finite (as in typical experiments, where $N\sim 10^3-10^5$), we find that fluctuations of the collective spin can be important.
\end{abstract}%
\maketitle

\section{Introduction}
Coherent states (CS) are an essential tool of modern physics. The original (or ``standard") coherent states of a harmonic oscillator are quasi-classical wavepackets following closely the classical oscillating trajectories with minimal uncertainty in their position and momentum. Mathematically, they are obtained by acting with a displacement operator $\hat{D}(\gamma)$ on the ground state, with $\gamma$ a complex number indexing the CS. This definition can be generalized to other systems, in particular if the Hilbert space $\mathcal{H}$ of the physical system under consideration is an irreducible representation space for a particular Lie group $\mathcal{G}=\{ \hat{G}({\bm \gamma}) \}_{\bm \gamma}$ indexed by a continuous label ${\bm \gamma}$. Following \cite{perelomov1977a,zhang1990a}, generalized CS are then obtained by acting with all elements of the group on some reference state $\vert {\rm ref} \rangle$ in $\mathcal{H}$. The operators $\hat{G}({\bm \gamma})$ generalize the displacement operators introduced above. The group structure ensures that the family of generalized CS generated in this way is closed. When the group is $SU(2)$ and the representation space the Fock space with $N$ particles in two modes (which describes, for instance, an ensemble of $N$ two-level atoms or spin $1/2$ particles), this construction leads to the well-known $SU(2)$ CS, sometimes simply called ``spin coherent states''. Spin CS are used extensively in fields as diverse as quantum optics \cite{arecchi1972a} or magnetism \cite{auerbach}. Similarly, one can introduce $SU(3)$ CS to describe the collective state of $N$ three-level atoms \cite{gitman1993a,cao1995a,gnutzmann1998a} or spin 1 lattice models of magnetic materials ({\emph e.g.} \cite{ivanov2003a,lauechli2006a}). 

In this article, we apply the $SU(3)$ CS formalism to the study of the equilibrium properties of a spin $1$ Bose-Einstein condensate at finite temperature, where the three modes are the three Zeeman states with magnetic quantum numbers $m=0,\pm 1$ along a given quantization axis ${\bm z}$. We assume the validity of the single-mode approximation (SMA), which considers that all atoms condense in the same spatial mode \cite{law1998a,ho1998a,ohmi1998a,ho2000a,koashi2000a,castin2001a,yi2002a,barnett2011a,lamacraft2011a}. Within the SMA, the resulting formalism allows one to describe an arbitrary collective spin state of $N$ bosons in the same spatial mode, and is sufficiently simple to yield explicit analytic predictions that can be used to analyze experimental results. We discuss in detail the expected thermal equilibrium for the situation of current experimental relevance, where $N$ is large and where the magnetization of the system is conserved \cite{stamperkurn2013a}. We illustrate the method by calculating the first moments of the $m=0$ population as well as its probability density function in various cases of experimental interest. In the thermodynamic limit $N\rightarrow \infty$, the theory reduces to the usual mean-field treatment \cite{ho1998a,ohmi1998a,zhang2003a}. When $N$ is large but finite (as in typical experiments, where $N\sim 10^3-10^5$), we find that fluctuations of the collective spin can be pronounced and experimentally observable. 

%\section{Basic formalism}
\section{$SU(3)$ coherent states}
In order to establish notations, we start by considering a single spin 1 particle. An arbitrary pure state $\vert {\bm \zeta} \rangle  = {\mathcal{U}} \vert m=+1 \rangle$ can be obtained by applying a $SU(3)$ transformation ${\mathcal{U}}$ to the maximally polarized state. We find convenient to express $\vert  {\bm \zeta} \rangle $ as
\begin{align}
\vert  {\bm \zeta} \rangle  =
\begin{pmatrix}
\sqrt{\frac{1-n_0+m_z}{2}} e^{i \frac{\Theta+\alpha}{2}}\\
\sqrt{ n_0}\\
\sqrt{\frac{1-n_0-m_z}{2}} e^{i \frac{\Theta-\alpha}{2}}
\end{pmatrix}.
\end{align}
The global phase is chosen such that the $m=0$ component is real. Physically, $n_0 \in [0,1]$ corresponds to the reduced population in $m=0$ and $m_z \in [-1,1]$ to the reduced longitudinal magnetization, both normalized to the total particle number ($0 \leq 1-n_0- \vert m_z \vert \leq 1$). We have also introduced two relative phases, $\Theta = \phi_{+1}+\phi_{-1}-2\phi_0 \in [0,2\pi]$ and $\alpha = \phi_{+1}-\phi_{-1}\in [0,4\pi]$, with $\phi_m$ the argument of the component $\zeta_m$. The phase $\alpha$ can be related to the orientation in the $x-y$ plane of the average transverse spin vector ${\bm s}_\perp=\langle \hat{s}_x\rangle {\bf e}_x+\langle \hat{s}_y\rangle {\bf e}_y$, and the phase $\Theta$ to the magnitude of ${\bm s}_\perp$, 
\begin{align}\label{eq:Sperp}
s_\perp^2 & = 2 n_0 \left(1-n_0+\sqrt{(1-n_0)^2- m_z^2}  \cos\Theta\right).
\end{align}

We now turn to the case of $N$ bosons. The Fock space for $N$ bosons with three possible internal states remains a representation space for $SU(3)$, and we can generate a family of states $\vert  {\bm \zeta}^N \rangle$ by a $SU(3)$ transformation acting on the maximally polarized state $ \vert N : m=+1 \rangle$ \cite{gnutzmann1998a}. The states $\vert {\bm \zeta}^N \rangle$, explicitly given by
\begin{align}\label{eq:SU3CS}
\vert  {\bm \zeta}^N \rangle & = \frac{1}{\sqrt{N!}} \left( {\bm \zeta} \cdot \hat{{\bm a}}^\dagger \right)^N \vert \varnothing \rangle,
\end{align}
in second quantized notation, describe an ensemble of $N$ bosons condensing simultaneously in the same spin state $\vert {\bm \zeta} \rangle$. Here $\hat{{\bm a}}=(\hat{a}_{+1},\hat{a}_{0},\hat{a}_{-1})^T$ is a vector notation for the annihilation operators in each Zeeman substate, and $\vert \varnothing \rangle$ is the vacuum state.

%\subsection{Main properties of generalized coherent states}
%\label{sec:CS}
The $SU(3)$ CS $\vert {\bm \zeta}^N \rangle$ are generalized CS in the sense described in the Introduction. They provide a resolution of the identity operator \cite{perelomov1977a,zhang1990a,yaffe1982a,gitman1993a,cao1995a,gnutzmann1998a}, $\int d{\bm \zeta}~\vert {\bm \zeta}^N \rangle \langle {\bm \zeta}^N \vert={\bm 1}$, with the measure $d{\bm \zeta}=(N+1)(N+2)/8\pi^2 \times dn_0 dm_z d\Theta d\alpha$. This implies that $SU(3)$ CS form a basis of Fock space, which is overcomplete since CS are not orthogonal to each other. Operators acting in Fock space can be represented as diagonal operators in the CS basis \cite{perelomov1977a,zhang1990a}. 

\subsection{Classical limit $N \rightarrow \infty$}
\label{sec:classical}
The main interest of using the basis of coherent states is the simplicity of the resulting theory in the large $N$ limit, as discussed in details by Yaffe \cite{yaffe1982a} (see also \cite{castin2001a} for a discussion focusing on Bose-Einstein condensates). In this limit, the scalar product $\langle {\bm \zeta}^N \vert {\bm \zeta'}^N\rangle = \left( {\bm \zeta} \cdot {\bm \zeta}'\right)^N$ becomes very peaked around ${\bm \zeta}={\bm \zeta}'$ due to the large $N$ power. This allows one to perform the approximation
\begin{align}\label{eq:scalarprod}
\langle {\bm \zeta}^N \vert {\bm \zeta'}^N\rangle & \approx \mathcal{A} ~\delta^{(N)}({\bm \zeta}-{\bm \zeta}'),
\end{align}
with $\mathcal{A}$ a normalization coefficient. The function $\delta^{(N)}$ is normalized to unity with the measure $d{\bm \zeta}$ and vanishes very quickly (on a typical scale $\sim 1/\sqrt{N}$ in each generalized coordinate) when ${\bm \zeta}'$ moves away from ${\bm \zeta}$. As a result, $\delta^{(N)}$ tends to a Dirac distribution $\delta({\bm \zeta}-{\bm \zeta}')$ when $N\rightarrow \infty$. The normalization of the CS implies the relation $ \int d{\bm \zeta}' \left[ \mathcal{A}\delta^{(N)}({\bm \zeta}-{\bm \zeta}') \right]^2=1$, which can be used to evaluate the normalization coefficient. 

In the large $N$ limit, the quasi-orthogonality between two CS expressed in Eq.\,(\ref{eq:scalarprod}) greatly simplifies the computation of expectation values. In the CS states basis, a $k-$body normally ordered operator $\hat{O}$ with $k \ll N$
can be approximated as
\begin{align}
\langle {\bm \zeta}^N \vert \hat{O} \vert {\bm \zeta'}^N \rangle
 \approx \langle {\bm \zeta}^N \vert \hat{O} \vert {\bm \zeta}^N\rangle \mathcal{A} ~\delta^{(N)}({\bm \zeta}-{\bm \zeta}')
\end{align} 
As a result, the expectation value in a CS of a product of two (few-body) operators obeys a simple rule, $ \langle {\bm \zeta}^N \vert  \hat{O}\hat{P} \vert {\bm \zeta}^N \rangle \approx  \langle {\bm \zeta}^N \vert  \hat{O}\vert {\bm \zeta}^N \rangle \langle {\bm \zeta}^N \vert \hat{P} \vert {\bm \zeta}^N \rangle$. This property allows one to compute thermodynamic averages in a thermal state described by a density operator $\hat{\rho}$ using the intuitively appealing formula,
\begin{align}
\langle \hat{O} \rangle \approx  \int d{\bm \zeta}~ \langle {\bm \zeta}^N  \vert \hat{O} \vert {\bm \zeta}^N\rangle \times \langle {\bm \zeta}^N  \vert  \hat{\rho} \vert {\bm \zeta}^N\rangle.
\end{align}
Effectively, $\langle {\bm \zeta}^N  \vert  \hat{\rho} \vert {\bm \zeta}^N\rangle$ plays the role of a classical distribution function in ${\bm \zeta}$ space. For the canonical ensemble, for instance, we have 
\begin{align}
\langle {\bm \zeta}^N \vert  \hat{\rho} \vert {\bm \zeta}^N \rangle & \approx  \frac{1}{\mathcal{Z}} e^{-\beta \langle {\bm \zeta}^N \vert\hat{H}\vert {\bm \zeta}^N \rangle},
\end{align}
with $\mathcal{Z}=\int d{\bm \zeta}e^{-\beta \langle {\bm \zeta}^N \vert\hat{H}\vert {\bm \zeta}^N \rangle}$ the partition function and with $\beta=1/k_B T$ the inverse temperature. 

In this article, we consider particle numbers $N$  which are large but finite, so that the orthogonality relation between two coherent states holds only approximately. In the following, we use systematically the large $N$ limit, which should thus be understood as the dominant power of $N$ in a $1/N$ expansion. 

\section{Application to the statistical mechanics of a spin 1 BEC with constrained magnetization}

%\subsection{Single-mode Hamiltonian}

We now apply this formalism to the description of the low temperature properties of a trapped gas of ultracold spin 1 bosons. As indicated in the Introduction, we assume that all atoms occupy the same spatial mode $\overline{\phi}({\bm r})$ (but not necessarily the same spin state) \cite{yi2002a}. The Hamiltonian describing the spin dynamics is then \cite{law1998a}
\begin{align}\label{eq:H}
\hat{H} = \frac{U_s}{2N} {\bm \hat S}^2 - q \hat{N}_{0},
\end{align}
where $U_s$ is the spin interaction energy per atom, $q>0$ is the quadratic Zeeman energy, $ {\bm \hat S}$ is the total spin operator, and $\hat{N}_{m}$ is the number operator for the Zeeman state $m$. 

%\subsection{Strongly constrained generalized statistical distribution}\label{sec:statensexact}

In most experiments with spinor gases ({\it e.g.} \cite{schmaljohann2004a,chang2004a,sadler2006a,liu2009a,guzman2011a,bookjans2011a,jacob2012a}), the system is prepared with a prior distribution that depends on the particular experimental sequence, noted $p_M$ ($M$ being the eigenvalues of $\hat{S}_z$). Typically, $p_M$ is peaked around the average value $\overline{M}_z=N\overline{m}_z$. Because $\hat{S}_z$ commutes with the interaction Hamiltonian, the prior distribution is essentially preserved by binary collisions driving the gas towards kinetic equilibrium.  In order to account for the experimental situation, we use a generalized statistical ensemble, where the energy and the populations $p_M$ are conserved on average. Maximizing the entropy following the standard Gibbs procedure leads to the density matrix
\begin{align}
\hat{\rho} & = \frac{1}{\mathcal{Z}} \sum_{M=-N}^{M=N} e^{-\mu_M} \hat{P}_M e^{-\beta \hat{H}} \hat{P}_M,
\end{align}
where $\hat{P}_M$ is the projector on the subspace $M$ and where the $\mu_M$'s are Lagrange multipliers introduced to enforce the conservation of the probabilities $p_M = \langle \hat{P}_M \rangle$.

Although this procedure would be the most rigorous one, it leads to a rather complicated formalism, where $N$ constants of motion are required to describe the ensemble. Instead of constraining the full distribution, we choose in the following to constrain only the first two moments $\overline{m}_z$ and $\Delta m_z^2 =\overline{m_z^2}-\overline{m}_z^2$. We expect that the differences from the more rigorous formalism will not be significant as long as only few-body observables are computed. Constraining the first two moments of $m_z$ leads to a density matrix
\begin{align}
\hat{\rho} & = \frac{1}{\mathcal{Z}} e^{-\beta \hat{H}- \lambda_1 \hat{S}_z - \lambda_2 \hat{S}_z^2}=\frac{1}{\mathcal{Z}} e^{-\beta \hat{K}},
\end{align}
where $\lambda_{1/2}$ are two Lagrange multipliers.

In the large $N$ limit, the partition function is determined by the free energy $K \equiv \langle  {\bm \zeta}^N \vert\hat{K}  \vert  {\bm \zeta}^N \rangle$. Using Eqs.\,(\ref{eq:Sperp},\ref{eq:H}) and the properties of CS, we rewrite $K$ as
\begin{align}\label{eq:K}
\beta K &= \frac{\beta_z'}{2}  \left( m_z -m_z^\ast \right)^2  - \eta n_0 \\
& + \beta' n_0 \left( 1-n_0+\sqrt{(1-n_0)^2-m_z^2}\cos(\Theta)\right).\nonumber
\end{align}
We have introduced two dimensionless parameters
\begin{align}
\eta=N\beta q,~~\beta'=N\beta U_s,
\end{align}
as well as two new Lagrange multipliers $\beta_z'=\beta' +2N^2\lambda_2$ and $m_z^\ast=-N\lambda_1/\beta_z'$ determined by the two constraints $\overline{m}_z =   \frac{1}{\mathcal{Z}} \int d{\bm \zeta} m_z e^{-\beta K}$ and $\Delta m_z^2 =   \frac{1}{\mathcal{Z}} \int d{\bm \zeta} \left(m_z-\overline{m}_z \right)^2 e^{-\beta K}$. The parameter $m_z^\ast$ is approximately equal to the average magnetization (but not exactly, unless $\overline{m}_z=0$). In the natural energy unit of $NU_s$, the parameter $\beta_z'$ can be interpreted as the inverse of a longitudinal pseudo-temperature characterizing the fluctuations of $\hat{S}_z$ in the prior distribution defining our generalized ensemble. In comparison, the inverse temperature $\beta'$ characterizes the fluctuations of the transverse components $\hat{S}_x,\hat{S}_y$. A purely thermal prior distribution of $m_z$ is characterized by $\beta_z'=\beta'$, a narrow prior distribution by $\beta_z' \gg \beta'$ and a broad one by $\beta_z' \ll \beta'$. The two dimensionless parameters $\eta$ and $\beta'$ allow us to specify the thermodynamic state completely given the two constraints on $\overline{m}_z$ and $\Delta m_z$. We emphasize that $\eta$ and $\beta'$ are both proportional to the total atom number $N$, which reflects the fact that we are dealing with fluctuations of a collective variable. 

In principle, the set of equations above can be used to characterize the collective thermodynamic state for any values of the parameters, $\overline{m}_z,\Delta m_z,\eta,\beta'$. In the following, we will illustrate the usefulness of the $SU(3)$ CS approach in particular regimes, where analytical results can be obtained : the mean field regime (where differences from the $T=0$ mean field theory are small), the regime of small $n_0$ for antiferromagnetic interactions, and the regime of strong spin fluctuations when $\overline{m}_z=0$. For simplicity, we restrict ourselves to the experimentally relevant case where the distribution of $m_z$ fulfills $\Delta m_z \ll 1$ ($\overline{m}_z$ can be chosen arbitrarily). To reduce the number of varying parameters, we consider a system at a fixed temperature $T$, atom number $N$ and interaction strength $U_s$ (so that $\beta'$ is constant). We vary the quadratic Zeeman energy (or equivalently $\eta$) and the average magnetization.

\section{Mean-field regime at $T \neq 0$}

The thermodynamics is controlled by the behavior of the free energy $K$. For all choices of the parameters $\overline{m}_z$ and $q$, $K$ has a well-defined minimum which depends on the sign of the spin-exchange interaction $U_s$. At $T=0$ the atoms condense in that minimum \cite{zhang2003a}, which determines the phase diagram \cite{zhang2003a,liu2009a,jacob2012a}. For $U_s<0$ (ferromagnetic interactions), the minimum is obtained for $\Theta=0$ and a certain value $n_0^\ast >0$ which maximizes the magnitude of the transverse spin. For $U_s>0$ (antiferromagnetic interactions), the minimum is obtained for $\Theta=\pi$ and the value $n_0^\ast$ which minimizes $S_\perp$. Differently from the ferromagnetic case, $n_0^\ast$ is zero until a critical value $q_c=U_s \left( 1-\sqrt{1-\overline{m}_z^2} \right)$ above which it becomes positive. 
%For very large $q$, the population in $m=0$ asymptotically tends to $n_0^\ast =1-\overline{m}_z$ irrespective of the sign of $U_s$. 

\begin{figure}
\begin{tabular}{lr}
\includegraphics{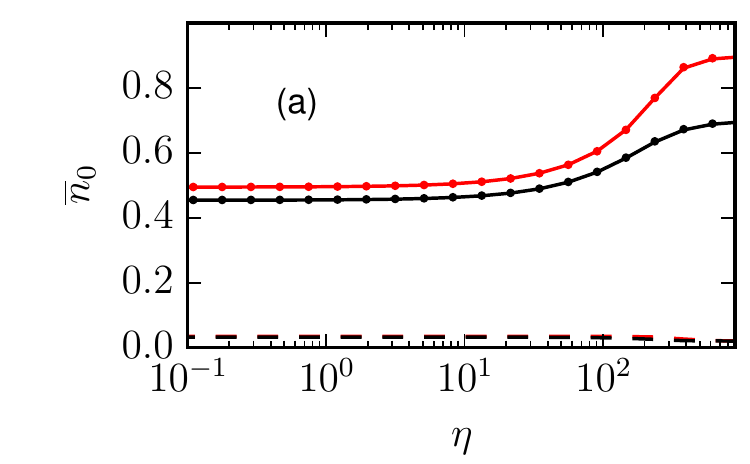}\\\includegraphics{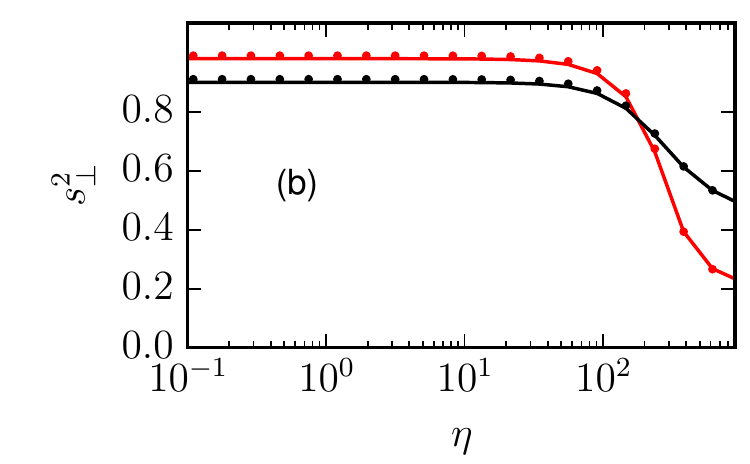}
\end{tabular}
\caption{{\bf (a)}: Moments of $n_0$ versus $\eta$ in the ferromagnetic case ($U_s<0$) for magnetizations $\overline{m}_z=0.1$ (red lines and symbols) and $\overline{m}_z=0.3$ (black lines and symbols). The solid lines show the average populations, the dashed lines the standard deviations, and the small dots the expected behavior at $T=0$. {\bf (b)}: magnitude of the transverse spin. The graphic conventions are the same as in {\bf (a)}. The values $\beta'=N\beta U_S=200$ and $\Delta m_z=0.02$ were used for all plots.}
\label{fig:results_ferro_constrained}
\end{figure}

One normally expects that for sufficiently low temperatures, the system only explores the vicinity of the minimum. The free energy then equals its value $K^\ast $ at the minimum plus small additional terms, corresponding to Gaussian fluctuations around the minimum with a typical spread $\Delta n_0,\Delta \Theta \sim 1/\beta'$ (up to coefficients depending on $q/U_s$ and $\overline{m}_z$). This describes well the case of ferromagnetic systems, where the finite $T$ solution is always close to the zero temperature one as shown in Figure~\ref{fig:results_ferro_constrained}. The small differences are due to the combined effects of fluctuations and of the spread of $m_z$, which are included in the finite $T$ calculation but not in the $T=0$ one. The Gaussian expansion is valid provided $\beta' \gg 1$, or
$k_B T \ll N U_s$. For typical experimental values in \cite{liu2009a,jacob2012a}, $N\sim 10^4$, $U_s/k_B\sim 2 ~$nK and $T \sim 100~$nK, $\beta' \sim 200$ is indeed large. This leads to results for the thermodynamic observables essentially identical to the ones obtained at zero temperatures, up to small corrections of magnitude $\sim 1/\beta'$. 

Antiferromagnetic systems behave differently (see Fig.~\ref{fig:results_aferro_constrained}). For $q<q_c$, the value of $\overline{n}_0$ is not zero and fluctuations are comparable to the mean value. This strongly differs from the conclusion drawn from the $T=0$ theory. Both effects become larger when $m_z \rightarrow 0$. 

\section{Antiferromagnetic systems with small $n_0$}\label{sec:smalln0}

\begin{figure}
\begin{tabular}{lr}
\includegraphics{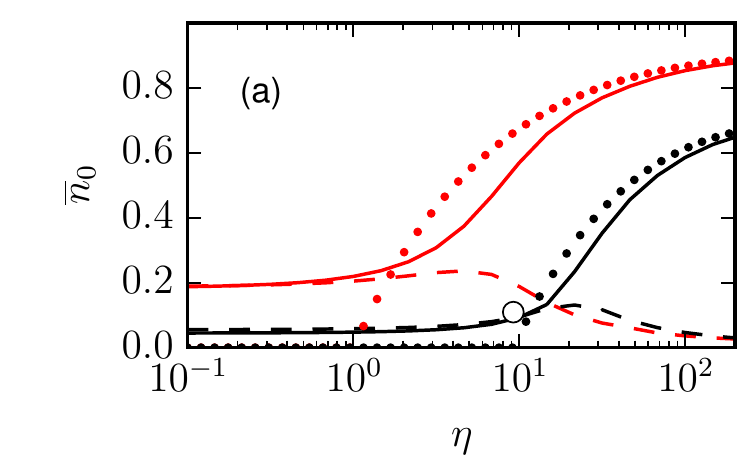}\\\includegraphics{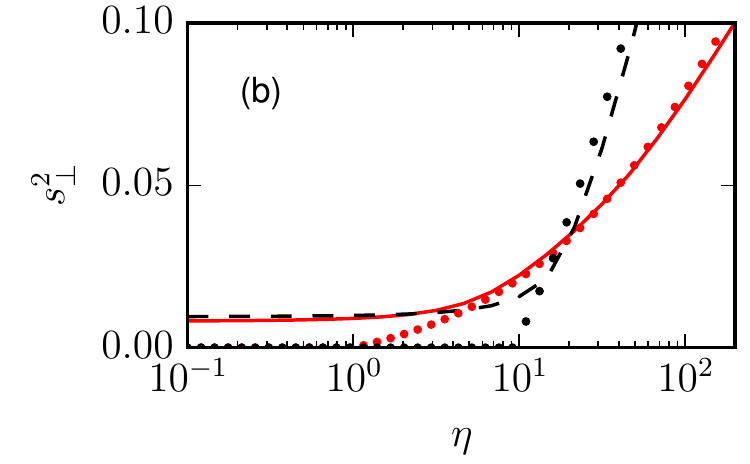}
\end{tabular}
\caption{{\bf (a)}: Moments of $n_0$ versus $\eta$ in the antiferromagnetic case ($U_s>0$) for magnetizations $\overline{m}_z=0.1$ (red lines and symbols) and $\overline{m}_z=0.3$ (black lines and symbols). The solid lines show the average populations, the dashed lines the standard deviations, and the small dots the expected behavior at $T=0$. The larger white dot shows the analytical limit for $q=q_c$, $\overline{n_0}\vert_{q=q_c} \approx 0.34(1-\overline{m}_z^2)^{1/4}/A^{1/2}$ for $\overline{m}_z=0.3$. {\bf (b)}: magnitude of the transverse spin. The graphic conventions are the same as in {\bf (a)}. The values $\beta'=N\beta U_S=200$ and $\Delta m_z=0.02$ were used for all plots.}
\label{fig:results_aferro_constrained}
\end{figure}

In order to understand the regime with $q<q_c$ and $T \neq 0$ better, we assume that $n_0$ remains small (which requires a finite magnetization $m_z$, see below) and expand the free energy around $\Theta=\pi$ and $n_0=0$. This gives (after integration over $m_z$ and $\Theta$) the partition function as ${\mathcal{Z'}}  \propto \int_0^1 dn_0 ~e^{-\beta K_{\rm eff}}/\sqrt{n_0}$, with an effective free energy
\begin{align}\label{eq:expK_smalln0}
\beta K_{\rm eff} &= B n_0^2-A \left(\frac{q}{q_c}-1 \right) n_0+\mathcal{O}\left(n_0^3\right),
\end{align}
with $A= N\beta q_c$ and $B=A\sqrt{1-\overline{m}_z^2}$. For $q>q_c$, the free energy has a minimum for $n_0^\ast \approx A(q/q_c-1)/2B >0$, where the first derivative of $K_{\rm eff}$ vanishes: We retrieve the mean field regime.  

The equilibrium population in $m=0$ is given by $\overline{n_0}\vert_{q=0}\approx1/A$ for $q=0$ and by $\overline{n_0}\vert_{q=q_c}  \approx a_1(1-\overline{m}_z^2)^{1/4}/A^{1/2}$ ($a_1 \approx 0.34$) for $q=q_c$. For high $\overline{m}_z$, we have $A \approx \beta' \gg 1$, and correspondingly small population in $m=0$. For small $\overline{m}_z$, we have $A \approx \beta' \overline{m}_z^2/2$ : the population in $m=0$ is small only if $\overline{m}_z \gtrsim 1/\sqrt{\beta'}$. Note that when this is not fulfilled ($A \sim 1$), the expansion in Eq.\,(\ref{eq:expK_smalln0}) is not valid. The average transverse spin per atom also becomes finite at $T \neq 0$. In the limit where $n_0$ remains small, we find from Eq.~(\ref{eq:Sperp})
\begin{align}
s_\perp^2 & \approx \frac{2 \overline{n_0} q_c}{U_s}+\mathcal{O}\left( n_0^2,n_0 \Theta^2\right).
\end{align}
In the limit $q \rightarrow 0$, we find $s_\perp^2\approx 2/\beta'$ independent of $\overline{m_z}$. For the typical values of $\beta'=200$ given above, $s_\perp^2$ reaches a few percent below $q_c$ (see Figure~\ref{fig:results_aferro_constrained}b), which is experimentally measurable. 

\section{Spin fragmentation for antiferromagnetic interactions and small magnetizations}

For antiferromagnetic interactions ($U_s>0$), a special situation occurs near $\overline{m}_z=0$ where the critical $q_c$ vanishes. In this regime, arbitrary large fluctuations of $n_0$ are possible, which makes the approximation used in the previous paragraph invalid. Going back to the spin-dependent free energy Eq.~(\ref{eq:K}), we expand around $\Theta=\pi$ and $m_z =0$, 
\begin{align}\label{eq:Knematic}
\beta K \approx \frac{1}{2} \frac{m_z^2}{\sigma^2}
+ \beta' n_0(1-n_0)\frac{\left(\Theta-\pi\right)^2}{2} -\eta n_0,
\end{align}
with 
\begin{align}
\sigma & = \sqrt{ \frac{1-n_0}{\beta' n_0+\beta_z'(1-n_0) }}.
\end{align}
The overall minimum, determined by the quadratic Zeeman energy [last term in Eq.~(\ref{eq:Knematic})], is at $n_0^\ast=1$, $\Theta= \pi$ and $m_z=0$. The curvature near the minimum along the $n_0$ direction vanishes. In this case, the condition $\beta' \gg 1$ is not sufficient to ensure the distribution is peaked around the mean-field solution: One must also have $\eta \gg 1$. When $q=0$, this is never fulfilled and instead of a single minimum, one finds instead a family of degenerate minima, corresponding to the so-called polar (or ``spin-nematic") states with vanishing spin \cite{ho1998a,ohmi1998a,castin2001a,desarlo2013a}. Because of the broad distribution, the system displays in the limit $\eta \ll1$ large fluctuations in the individual populations $n_0, n_{+1},n_{-1}$. As discussed in Refs.~\cite{ho2000a,castin2001a,koashi2000a,mueller2006a,barnett2011a,desarlo2013a}, this is a signature for fragmentation of the condensate which can occupy any of the quasi-degenerate states or an arbitrary superposition of them. We stress again that these fluctuations are a mesoscopic effect, and disappear in the thermodynamic limit where they are confined to a vanishingly small window around $\eta=0$. For $\beta_z'=\beta'$, we recover the previous results obtained at finite temperatures \cite{desarlo2013a}. The present theory is able to go further by accounting for the most general situation where $\beta_z' \neq \beta'$. 

\begin{figure}
\includegraphics{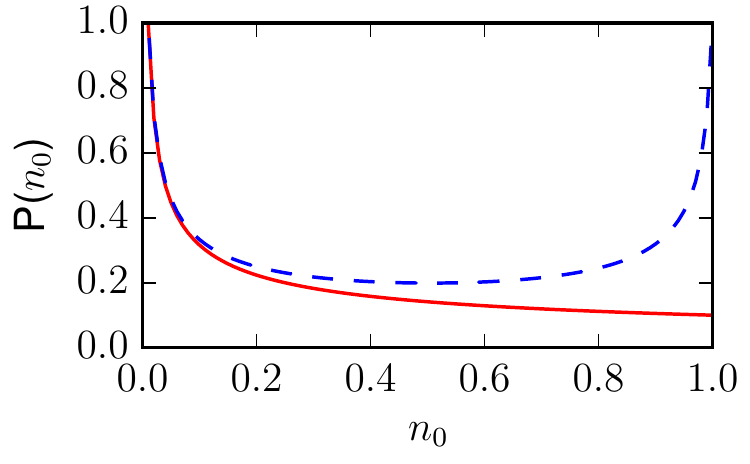}
\caption{Probability density $P(n_0)$ for the reduced population in $m=0$ (the densities are not normalized). The red solid line shows the distribution for $\beta_z'=\beta'$ and the blue dashed line for $\beta_z'=100 \beta'$. }
\label{fig:Pn0}
\end{figure}

After integration over $\Theta,m_z$, we obtain the (unnormalized) marginal distribution function of $n_0$ in a simple form,
\begin{align}
P(n_0)& \propto \frac{e^{\eta n_0} } {\sqrt{n_0( \beta' n_0+\beta_z'(1-n_0)) }}.
\end{align}
Other marginal distributions ({\it e.g.} for $m_z$) could be obtained in a similar way. For $q=0$ (see Figure~\ref{fig:Pn0}), the distribution of $n_0$ changes from an asymmetric characteristic square-root shape, $P(n_0) \propto 1/\sqrt{n_0}$ \cite{tasaki2013a}, to a symmetric shape $P(n_0) \propto 1/\sqrt{n_0(1-n_0)}$ when $\beta_z'$ changes from $1$ (unconstrained prior distribution of $m_z$) to $+\infty$ (narrow prior distribution of $m_z$). For small $\eta\ll 1$, the distribution $P(n_0)$ is always broad, so that the qualitative conclusions about spin fragmentation and large population fluctuations are unchanged. %These changes can be verified experimentally.

\begin{figure}
\centering
\includegraphics{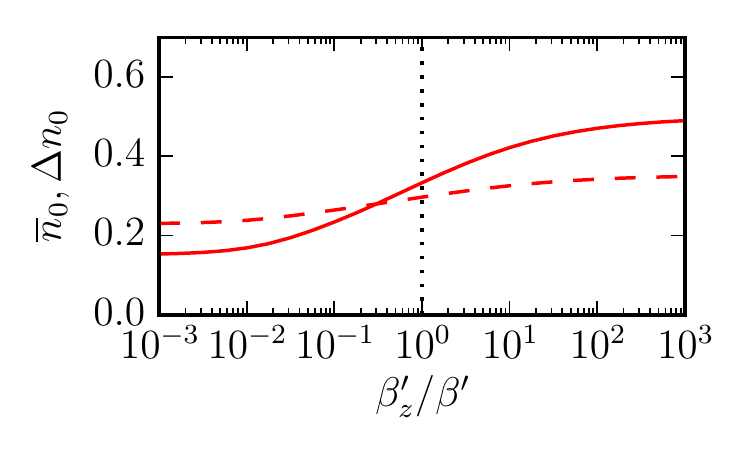}
\caption{Approximate theory in the case $\overline{m}_z=0$ and $q=0$: Average  (solid line) and standard deviation (dashed line) of the population $n_0$. The vertical dotted line marks the unconstrained case ($\beta_z'=\beta'$). }
\label{fig:aferro_eta0_m0}
\end{figure}

To discuss the influence of the prior distribution of $m_z$, we set $q=0$ and plot in Figure~\ref{fig:aferro_eta0_m0} the first two moments of $n_0$ versus $\beta_z'$. For $\beta_z'=1$, which corresponds to the situation without constraint where $\overline{n}_0 =1/3$, we find $\Delta n_0\approx 0.30$ and $\Delta m_z =2/(3\beta')$. With increasing $\beta_z'$, the prior distribution of $m_z$ becomes narrower: $\overline{n}_0$ goes from $1/3$ to $1/2$ and $\Delta n_0$ increases slightly \cite{desarlo2013a}. Conversely, $\beta_z'<1$ corresponds to a prior distribution of $m_z$ broader than the one without constraint : the average $\overline{n}_0$ decreases below $1/3$ (the fluctuations of $n_0$ decrease as well). 

\section{Conclusion}
In conclusion, we have proposed a description of the collective equilibrium state of a spin 1 Bose-Einstein condensate based on $SU(3)$ coherent states. Using this formalism, the conservation of the magnetization is accounted for by introducing a prior distribution for $m_z$ and constructing a generalized statistical ensemble that preserves its first moments. We have computed moments of various quantities, and their probability distribution function (for example, for the population in the $m=0$ Zeeman substate), that can be directly compared to experiments. Going beyond thermodynamics as studied in this paper, we expect, in analogy with what has been done with $SU(2)$ coherent states, that the $SU(3)$ coherent state formalism can be used to study the collective dynamics leading to spin oscillations \cite{chang2005a,kronjaeger2005a,black2007a,davidson2015a} or spin-nematic squeezing \cite{gross2011a,luecke2011a,hamley2012a,yukawa2013a}. Another interesting direction is to extend the formalism to larger groups $SU(N)$ with $N >3$ to describe condensates with higher spin. This is relevant for instance for experiments with Chromium atoms with spin 3 \cite{pasquiou2011a,pasquiou2012a}, where spin exchange and magnetic dipole-dipole interactions both play an important role. 

\acknowledgments
We thank the members of LKB for stimulating discussions. We acknowledge support from IFRAF, from DARPA (OLE program), from the Hamburg Center for Ultrafast Imaging and from the ERC (Synergy grant UQUAM).

%
%\bibliographystyle{apsrev}
%\bibliography{BiblioSU3}

%
%
\end{document}